\documentstyle[art12]{article}
\begin{document}

\hspace*{12cm} June 1994

\vspace*{1.5cm}  \begin{center} \Large \bf
 Current-like variables \\
 in massive and massless integrable models\footnote{\footnotesize Lectures
delivered
 at the International School of Physics "Enrico Fermi", held in Villa
 Monastero, Varenna, Italy, from 28 June to 8 July 1994} \end{center}
\vspace{0.7cm} \begin{center}
 \large \bf L.Faddeev  \end{center}

\vspace*{1cm}  \begin{tabular}{l}
\it St.Petersburg Branch of the Steklov Mathematical Institute \\
\it Fontanka 27, St.Petersburg 191011, Russia  \\
 {\rm and} \\
 \it Research Institute for Theoretical Physics  \\
 \it P.O. Box 9 (Siltavuorenpenger 20C), SF-00014
 University of Helsinki, Finland
 \end{tabular}

 \vspace*{2cm}

\section{Introduction}

 In these lectures I shall deal with the representative examples of integrable
models of quantum field theory in $1+1$ dimensional space-time. The term
"integrable" first used in this domain  in \cite{1}, means that the equations
of
motion have enough conservation laws to be exactly soluble. In quantum case
the spectrum and eigenvectors of the Hamiltonian should be calculable. Twenty
years of development showed that it is indeed the case and the famous Bethe
Ansatz \cite{2} found its reincarnation as a main tool to make it possible
\cite{3}.
A parallel development was done in the 2-dimensional lattice classical
statistical mechanics \cite{4}.

 While the latter development was highly appreciated by the physics community,
the former was considered rather peripherical because of its main limitation -
space-time being two-dimensional. String theory \cite{5} changed the
situation,
when the two-dimensional world-sheet aquired a bona fide physical
interpretation. However it were the conformal (massless) examples of field
theoretical models \cite{6}, which become prominent in this connection. The
known
(massive) examples of integrable models were ressurected as a perturbation
(deformation) of integrable models \cite{7} and exist now in the minds of
young
generation in this guise.

 For me as a specialist working on integrable models for more than twenty
years the opposite picture is more natural : the conformal field theory
models are contractions of integrable models. One can construct and classify
integrable models
(at least classically \cite{8}) without any reference to their conformal
contractions. The latter are to appear afterwards as particular massless
limits.

 In these lectures I shall discuss some technique and results pertinent to
the ideology, underlined above. It is natural to treat the massive and
massless models on some common ground. The massive models have divergences and
must be regularized. One of the nice features of their theory is that one can
devise the lattice formulation where the integrability remains intact. Thus we
should deal with the lattice models. On the other hand, the massless models,
having chiral excitations (left and (or) right movers), use the current-like
dynamical variables. Thus, one is to devise the way of using the current-like
variables for the massive models instead of the traditional canonical pairs.
Both goals will be accomplished in these lectures. Finally, we shall work with
finite volume space, which is natural for the conformal field models.

\section{Dynamical variables}

 Lattice and finite volume regularizations lead to consideration of quantum
mechanical system with finite number of degrees of freedom. As basic
generators of the algebra of observables we shall take, following \cite{9},
\cite{10},
the set of dynamical variables $w_n , \;\;\; n=1,..,N$, subject to the
commutation relations

\[      w_n w_{n+1} = q^2 w_{n+1} w_n  , \]
\[ w_n w_m = w_m w_n \;,\;\;\;\;\;\;\;\;\;\;\;\; | n - m | \ge 2 , \]
and the periodicity condition
\[     w_{n+N} = w_n . \]
Here $q$ is a complex number, plaing the role of coupling constant. In most
of applications $q$ takes values on the circle
\[          | {q} | = 1  ,    \]
or
\[      q = e^{i \gamma \hbar} ,   \]
where $\gamma$ is real and we introduced explicetely the Planck constant
$\hbar$. This requirement is compatible with the unitarity of $w_n$ :
\[      w_n^{\star} = w_n^{-1} .  \]
However, from time to time it will be convenient to relax this condition.

 Algebra $\cal A$, generated by $w_n$, has one central element
\[        C = w_1 w_2 ... w_N     \]
for $N$ odd and two central elements
\[  C_1 = w_1 w_3 ... w_{N-1} , \;\;\;\;\;\;\;\;\;  C_2 = w_2 w_4 ... w_{N} \]
for $N$ even. Thus we have $\frac{N-1}{2}$ degrees of freedom for $N$ odd
and $ \frac{N}{2} - 1$ for $N$ even. In that follows we shall consider $N$
to be even ($N=2M$), so that we shall deal with $M-1$ degrees of freedom.

We can realize the generators $w_n$ in the form
\[           w_n = e^{i p_n} ,  \]
if the $p_n$ have the following commutation relations
\[  [ p_n, p_m ] = 2 i \hbar \gamma ({\delta}_{n+1,m} - {\delta}_{n,m+1} ). \]
However, the substitution
\[     w_n \rightarrow p_n = \frac{1}{i} \ln{w_n}  \]
is not harmless; in particular, variables $p_n$ belong not to the algebra
$\cal A$, but rather to its extension. For evident reasons we shall refer to
algebra $\cal A$, generated by $w_n$, as a compact one and to larger
algebra $\hat{\cal A}$, generated by $p_n$, as a noncompact.

 The algebra $\hat{\cal A}$ is much larger than $\cal A$. However, one can use
$w$-like operators to describe it. For this goal we shall introduce the
variables :
\[ \hat{w}_n = e^{i \frac{\pi}{\gamma} p_n} = {w_n}^{\frac{\pi}{\gamma}} . \]

 It is evident that the variables $w_n$ and $\hat{w}_n$ commute
 \[ \hat{w}_n w_m = w_m \hat{w}_n .  \]

 It is a less trivial statement, that $w$ and $\hat{w}$ together generate all
$\hat{\cal A}$ for any $q$ not being a root of unity. Algebra $\cal A$
then is a factor inside of $\hat{\cal A}$.

 To make these statements more exact one has to introduce some topology, which
is standard (at least for $ |q| = 1$), and use the language of v-Neumann
algebras. I do not have time to do it in these lectures. One thing is
remarkable and worth to be mentioned explicetely. The algebra $\cal A$ has a
trace at least for $C_1 = C_2 = 1$! If a generic element is given by
\[  a = \sum a_{i_1 ... i_{N-2}} w_{1}^{i_1} ... w_{N-2}^{i_{N-2}}  ,  \]
(we do not use the dependent variables $w_{N-1}, w_{N-2}$),
then
\[ {\rm tr} \; a = a_{0,0 ... 0} \;  .   \]
Using the commutation relations one can show that the main propertiy of trace
\[ {\rm tr} ( a b ) = {\rm tr} ( b a )   \]
holds. The trace of unity is equal to $1$, so we have here the example of a
finite factor and in general factor ${\rm II}_1$.

 The complimentary variables $\hat{w}_n$ have the same current-like
commutation relations
\[ \hat{w_{n+1}} \hat{w}_n = {\hat{q}}^2 \hat{w}_n \hat{w}_{n+1} ,   \]
where
\[ {\hat{q}} = q^{ - \frac{{\pi}^2}{{\gamma}^2} }  .                 \]
The role of the modular map
\[ \frac{\gamma}{\pi} \rightarrow - \frac{\pi}{\gamma}    \]
entering here will be commented on later.

 The described realisation of $w$ via $p$ is appropriate for the massless
case. Indeed, naive classical ($\hbar \rightarrow 0 $) and continuos limit,
realized via rescaling
\[ p_n = 2 \Delta p(x) \; , \;\;\;\;\; x = n \Delta  ,  \]
with $\Delta \rightarrow 0$ leads to the Poisson brackets
\[ \{ p(x) , p(y) \} = \gamma {\delta}^{\prime} (x - y)    \]
typical for current-like variables.

 However, the same algebra can be realized via canonical pairs $(\phi,\pi)$
with the commutation relations
\[ [{\phi}_n,{\pi}_m ] = - i \hbar \gamma \delta_{n m}, \;\;\;\; n,m = 1,...,M
\]
for example as
\[ w_{2n} = e^{2i \gamma \phi_n}, \;\;\;\;\; w_{2n+1} = e^{i(\pi_n -
\pi_{n-1})}  \;\; . \]
Here the continuos limit exists for
\[ \phi (x) = \phi_n , \;\;\;\; \Delta \pi (x) = \pi_n, \;\;\;\; x = n \Delta .
\]
The picture for space is

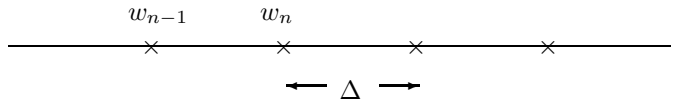
\begin{picture}(350,70)(0,0)
 \put(50,35){\line(1,0){250}}
 \multiput(100,32)(50,0){4}{$\times$}
 \put(95,45){$w_{n-1}$}  \put(145,45){$w_n$}
 \put(175,15){$\Delta$}
 \put(170,20){\vector(-1,0){15}}
 \put(190,20){\vector(1,0){15}}

\end{picture}  \\
for the first case and

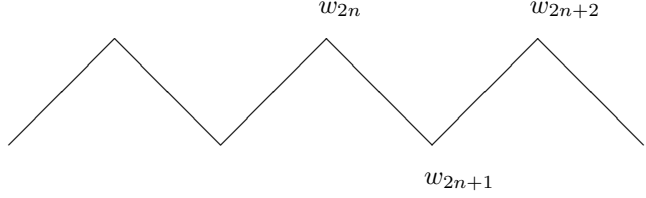
\begin{picture}(350,100)(0,0)
\multiput(50,20)(80,0){3}{\line(1,1){40}}
\multiput(90,60)(80,0){3}{\line(1,-1){40}}
\put(167,70){$w_{2n}$}
\put(247,70){$w_{2n+2}$}
\put(207,5){$w_{2n+1}$}
\end{picture}   \\
for  the second one. The first is to be used in massless case, while the
second
one for the massive case.

 Together with the current-like variables one can use also the vertex-like
ones.
In the continuos case they are defined by
\[   p(x) = {\psi}^{\prime} (x) .   \]
On a lattice massless and massive cases are to be treated separately. We shall
need only the second case. The analogue of space derivative is the ansatz ;
\[ w_n = \frac{\psi_{n+1}}{\psi_{n-1}}  .  \]
The inverse map $w \rightarrow \psi $ involves the product
\[ \psi_n = \prod\limits_{k \le n} \; w_k , \]
where $k$ and $n$ have an opposite parity.

 The last formula needs some clarification with respect to periodicity.
Instead we shall introduce the algebra with generators $\psi_n$
independentely. The whole algebra (denoted as $\cal B$) is generated by
$\psi_n , \;\;\; n = 1,...,2 M, \;\; C_1$ and $C_2 $ with relations
\[ \psi_n \psi_m = q \psi_m \psi_n \]
for $(n-m)$ odd; the quasiperiodisity holds as
\[ \psi_{2n + 2M} = C_1 \psi_{2n}      , \]
\[ \psi_{2n + 1 + 2M} = C_2 \psi_{2n + 1}      , \]
finally
\[ \psi_{odd} C_1 = q^2 C_1 \psi_{odd}  , \]
\[ \psi_{even} C_2 = q^2 C_2 \psi_{even}  , \]
and $\psi_{even}$ ($\psi_{odd}$) comute with $C_1$ ($C_2$).
This algebra has no central elements, so the number of degrees of freedom is
equal to $M+1$. It may be reduced to the algebra $\cal A$, generated by $w_n$,
if we introduced two constraints, fixing the values of (commuting) generators
$C_1$ and $C_2$.

 One can construct almost canonical basis, using some mixture of $w$ and
$\psi$ generators. Indeed, pairs $(w_{even},\psi_{even})$ or
$(w_{odd},\psi_{odd})$ are Weyl pairs, i.e.
\[ w_{2n} \psi_{2n} = q^2 \psi_{2n} w_{2n} , \;\;\; n = 1,...,M , \]
\[ w_{2n} \psi_{2m} = \psi_{2m} w_{2n} , \;\;\; n \neq m \]
and give us the system with $M$ degrees of freedom. Here the variable $C_1$
enters via the quasiperiodic conditions for $\psi_{even}$. The same is
true for $(w_{odd},\psi_{odd}, C_2)$. The
corresponding algebras will be denoted ${\cal C}_{even}$ and ${\cal C}_{odd}$.

 Now we are ready to consider concrete models of quantum field theory.

\section{ Massless case}

 The typical classical continuos equations of motion are first oder in time :
the free motion (left mover)
\[ \dot{p} (x) + p^{\prime} (x) = 0    \]
or KdV equation
\[ \dot{p} (x) + p (x) p^{\prime} (x) + p^{\prime \prime \prime} (x) = 0 \]
are the typical examples. They are produced by the Poisson stucture given above
and
the Hamiltonians
\[ H_0 = \frac{1}{2 \gamma} \int p^2 (x) dx ,  \]
\[ H_1 = \frac{1}{2 \gamma} \int ( \frac{1}{3} p^3 (x) + (p^{\prime} (x))^2 )dx
  \]
correspondingly. Here we shall treat only the first example. In spite of its
triviality it is known to be basic for CFT.

 On the rectangular space-time lattice

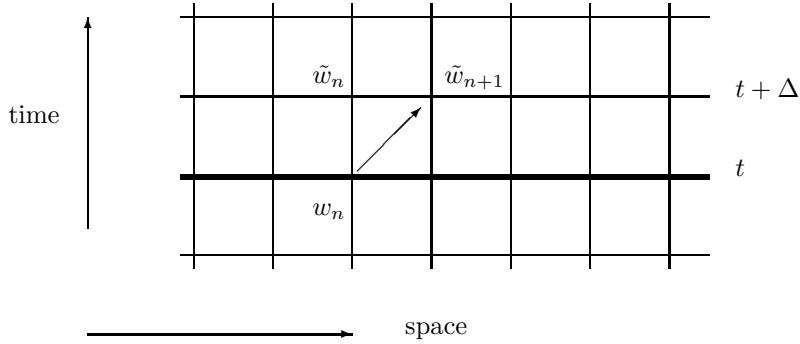
\begin{picture}(350,150)(10,0)
 \multiput(65,40)(0,30){4}{\line(1,0){200}}
 \multiput(70,35)(30,0){7}{\line(0,1){100}}
  {\thicklines \multiput(65,69)(0,1){2}{\line(1,0){200}}   }
 \put(132,72){\vector(1,1){24}}

 \put(30,50){\vector(0,1){80}}   \put(0,90){\rm time}
 \put(30,10){\vector(1,0){100}}   \put(150,10){\rm space}
  \put(275,70){$t$}   \put(275,100){$t+ \Delta$}
  \put(115,105){$\tilde{w}_n$}     \put(165,105){$\tilde{w}_{n+1}$}
 \put(115,55){$w_n$}
\end{picture}  \\
the simplest generalisation of equation of motion is
\[ w_n (t + \Delta) = \tilde{w}_n (t) = w_{n-1} (t) . \]
The map (simply a shift)
\[ w_n \rightarrow w_{n+1}  \]
is evidentely the automorphism of algebra $\cal A$. The integrability means
that we shall be able to construct it explicetely.

 We are looking for the operator $U$ such that
\[ w_{n+1} = U w_n U^{-1}  .  \]
Suppose that $U$ is multiplicatively local
\[ U = r(w_2) ... r(w_N).   \]
For $n = 1,2$ or $N$ we get
\[ w_{n+1} U = U w_n  \]
if
\[ w_{n+1} r(w_{n}) r(w_{n+1}) = r(w_{n}) r(w_{n+1}) w_{n}  .  \]
The basic commutation relations allows us to rewrite this relation as
follows
\[ r(q^{-2} w_{n}) w_{n+1} r(w_{n+1}) = r(w_{n}) w_{n}  r(q^{-2} w_{n+1}) \]
and separation of variables leades to the functional equation for the
function $r(w)$ :
\[ w \; r(w) = c \; r(q^{-2} w)  ,  \]
where $c$ is a constant. We shall normalize this equation as follows
\[
 \frac{r(q w)}{r(q^{-1} w)} = \frac{1}{w}  .
\]

 Now one can check that the operator $U$ serves as a shift also for the
variables $w_1,w_2$ and $w_N$ if the necessary condition
\[ C_1 = C_2  \]
is satisfied.

 The equation
\[ w_1 U = U w_N \]
leades to
\[ r(w_1) U = U r(w_N)  ,  \]
so that $U$ can be rewritten as
\[ U =  r(w_1) ... r(w_{N-1})   .   \]
This shows that the expression for $U$ is compatible with the periodicity; it
does not depend on a lattice site, we begin the multiplication of $r(w)$ with.
The only important things are the order of the factors and that the number of
them is equal to $N-1$ (not $N$ as one could superficially think).

 The solution of the functional equation inside the algebra $\cal A$ (i.e. in
power series
in $w$) gives an expression for $r$ as a $\theta$-function
\[ r(w) = \sum\limits_{k=-\infty}^{\infty} q^{k^2} w^k  ,  \]
which make sence only if $|q| < 1$. So, the unitary case is nonaccessible
for the generic $q$ with values on the unit circle. However, for $q$ being
odd root of unity
\[  q^{2Q+1} = 1 \]
the series
\[ r_Q (w) = \sum\limits_{k=-Q}^{Q}  q^{k^2} w^k  \]
solves the functional equation.

 The solution in the noncompact algebra $\hat{\cal A}$ exists for all $q$
and is rather simple :
\[ \hat{r} (p) = \exp{  \frac{i}{4 \gamma} p^2 } .  \]

  One immediately sees that in classical continuos limit the evolution
operator $U$
\[  U = e^{- i H \Delta}   \]
leads to the clasical expression for the Hamiltonian $H_0$.

 Thus we face the alternative :

$1).$ To work in algebra $\cal A$ and to bound ourselves in the unitary
situation with the $q$ being root of unity. The corresponding dynamical
system is finite dimensional, but the dependence on the coupling constant
$\gamma$ is rather erratic.

$2).$ To extend the algebra of observables to $\hat{\cal A}$. The evolution
operator is well defined for all real $\gamma$ and depends on $\gamma$
smoothly. However, the algebra of observables is infinite.

 More on this alternative will appear soon.

 The relevance of our model to CFT becomes explicite in classical continuous
case, when one introduces a set of generators of Virasoro algebra
\[ S(x) = p^2 (x) + p^{\prime} (x)    .\]
The Poisson brackets
\[ \{ S(x),S(y) \} = 2 \gamma (S(x) + S(y)) \delta^{\prime} (x-y) + \gamma
\delta^{\prime \prime \prime} (x-y)  \]
known for a long time as a second structure for the KdV equation \cite{11},
coinside with those of the Virasoro algebra \cite{12}. The Hamiltonian $H_0$
can
be written now as
\[ H_0 = \frac{1}{2 \gamma} \int S(x) dx  \]

 The natural question is if one can introduce a lattice analogue of these
formulas. The answer goes as follows \cite{9}.

 The function $r(w)$ admits the factorization
\[ r(w) = S(w) S(\frac{1}{w}) , \]
where $S(w)$ satisfies the following functional equation
\[ \frac{S(q w)}{S(q^{-1} w)} = \frac{1}{1+w}   . \]

 The solution in algebra $\cal A$ (so that $|q| < 1$) is given by the series
\[ S(w) = 1 + \sum^{\infty}_{k=1}  q^{\frac{k(k-1)}{2}} \frac{1}{(q^{-1} -q)
\ldots (q^{-k} - q^k)} w^k   , \]
or by the infinite product
\[ S(w) = \prod\limits_{k=1}^{\infty} (1 + q^{2 k + 1} w)   ,\]
or by the expression
\[ S(w) = \exp \sum^{\infty}_{k=1} \frac{(-1)^k w^k}{k(q^k - q^{-k})}
.\]
This function is well known in combinatorics as a $q-\Gamma$-function. The
first expression allowes also to call it $q$-exponent, whereas the last one
prompts the definition of $q$-dilogarithm.

 In the first guise $S(w)$ satisfies the $q$-exponential functional equation :
let $u$ and $v$ constitute a Weyl pair
\[   u v = q^2 v u   ,  \]
then
\[  S(u) S(v) = S(u+v)  . \]
I was not able to trace this result to its origine. Apparentely it was known
in combinatorics long ago. However, its combination with the functional
equation leads to the alternative equation
\[   S(v) S(u) = S(u+v+q^{-1} u v) = S(u) S(q^{-1} u v) S(v) , \]
which apparentely was mentioned first in \cite{9}. The first form of it will
be
used immediately. The second was shown in \cite{13} to be a quantum and
$q$-analogue
of the famous pentagon equation for the dilogarithm.

 Using the equation
\[ S(w^{-1}_n) S(w_{n+1}) = S(w^{-1}_n + w_{n+1} q^{-1} w_{n+1} w^{-1}_n), \]
we can easily rewrite the evolution operator $U$ in the form
\[ U =  S(w_2) S(w_2^{-1}) S(w_3) ... S(w_N) S(w_N^{-1}) = S(w^{-1}_1)
U(S(w^{-1}_N))^{-1} = \]
\[ = S(w^{-1}_1) S(w_2) S(w_2^{-1}) S(w_3) ... S(w_{N-1}^{-1}) S(w_N) = S(s_2)
... S(s_N), \]
where
\[ s_n =w^{-1}_{n-1} + w_n + q^{-1} w^{-1}_{n-1} w_n . \]
It was shown in \cite{14},\cite{15} that the variables $s_n$ are the natural
lattice
analogues of the $S(x)$. So, we shall call them the generators of the lattice
Virasoro algebra. To avoid missinterpritation let us stress that there is
no $q$-deformation here, the deformation parameter is only the lattice
spacing $\Delta$, which appears in the definition of the classical continuous
limit. In particular for $\Delta \rightarrow 0$, $\hbar \rightarrow 0$ we
have
\[ \frac{1}{4} (1 + s_n) = (1- {\Delta}^2 S(x)) + \cdots  .  \]
The last expression for the evolution operator $U$ is the analogue of the
classical expression for $H_0$ via $S(x)$.

 Now let us turn to the noncompact way. The factorisation
 \[ \hat r (p) = \hat S (p) \hat S (-p)  \]
can be achieved via the solution of the functional equation
\[ \frac{\hat S (p+\gamma)}{\hat S (p-\gamma)} = \frac{1}{1+e^{ip}}  \]
with the solution
\[ \hat S (p) = \\
 \exp \frac{1}{4} \int_{-\infty}^{\infty} \frac{e^{p x}}{\sinh{(\pi x)}
\sinh{(\gamma x)}} \frac{dx}{x}   , \]
where the singularity of the integral at $x=0$ is put below the contour of
integration. The integral is well defined for $|p| \le \pi + \gamma$ and then
$S(p)$ is continued for all real $p$ by means of the functional equation.

 Evaluating the integral through the residues we get the following expression
\[ \hat S (p) = S_q (w) S^{-1}_{\hat q} (\hat w)     ,  \]
where in the RHS we introduced explicetely dependence of $S(w)$ on $q$.
Whereas each separate factor does not make sence for $|q| = 1$, their ratio
does and smoothly dependes on $\gamma$. We see how two factors, constituting
the algebra $\hat{\cal A}$, harmoniously regularize each other.

 It is worthwhile to mention that the pentagon equation in the form
\[ \hat S (q) \hat S (p) = \hat S (p) \hat S (p+q) \hat S (q)  , \]
where
 \[ [p,q] = - i  \hbar \gamma    \]
stays intact in $\hat{\cal A}$.

  Two approaches recoincile if we note that the evolution operator $U$,
calculated by the second (noncompact) way, is an outer automorphism of the
(compact) algebra $\cal A$, reducing to the inner one as soon as it possible
(i.e. for $q$ being root of unity). In this sence the second way is more
general.

 The real difference appeares when we come to the spectral theory. Simple
spectrum of algebra $\cal A$ aquires the multiplicity in $\hat{\cal  A}$. So,
the evaluation of spectral characteristics such as partition function
depends drastically on the choice of algebra. In particular, in the second
way the coupling constants $\gamma$ and $\frac{{\pi}^2}{\gamma}$ appear
symmetrically. For me it is the origine of all fenomena called duality or
mirror symmetry, the appearence of the second quantum group the V. Kac table
for minimal models, the duality of the Sine-Gordon and massive Thirring
models, etc. This conviction still needs more work for its
complete vindication. On this note I shall finish the consideration of the
massless example and turn to the massive one.

\vspace*{1.5cm}

\section{Massive case}

The most famous example of the massive model is the Sine-Gordon equation
\[ \ddot{\phi}(x,t) -\phi^{\prime \prime}(x,t) + m^2 \sin{\phi }= 0 \]
for the scalar field $\phi(x,t)$ in two-dimensional space-time. In spite of
a vast literature devoted to this model (see references on history and results
in
\cite{8}), it still continues to reveal its new features. Here I shall show,
how the current-like variables can be used to treat the quantum lattice
variant of it.

As was already indicated above, the corresponding lattice is to be
light-like

\begin{picture}(350,120)(0,0)
  \multiput(80,60)(40,0){6}{\line(-1,-1){35}}
  \multiput(80,60)(40,0){6}{\line(1,1){45}}
  \multiput(80,60)(40,0){6}{\line(-1,1){45}}
  \multiput(80,60)(40,0){6}{\line(1,-1){35}}
 \put(60,10){\line(0,1){100}}
 \put(300,10){\line(0,1){100}}

 \put(62,35){$w_1$}       \put(77,65){$w_2$}
 \put(197,65){$w_{2n}$}   \put(217,32){$w_{2n+1}$}
 \put(277,65){$w_{2M}$}   \put(305,35){$w_1$}

 \multiput(80,59)(0,2){2}{\line(-1,-1){20}}
 \multiput(80,59)(0,2){2}{\line(1,-1){20}}

 \multiput(120,59)(0,2){2}{\line(-1,-1){20}}
 \multiput(120,59)(0,2){2}{\line(1,-1){20}}

 \multiput(160,59)(0,2){2}{\line(-1,-1){20}}
 \multiput(160,59)(0,2){2}{\line(1,-1){20}}

 \multiput(200,59)(0,2){2}{\line(-1,-1){20}}
 \multiput(200,59)(0,2){2}{\line(1,-1){20}}

 \multiput(240,59)(0,2){2}{\line(-1,-1){20}}
 \multiput(240,59)(0,2){2}{\line(1,-1){20}}

 \multiput(280,59)(0,2){2}{\line(-1,-1){20}}
 \multiput(280,59)(0,2){2}{\line(1,-1){20}}

\end{picture}   \\
with the dynamical variables $w_n$ given on the
"initial saw" (the vertical lines symbolize the periodic boundary condition
$w_n \equiv w_{n+2M}$).

Consider the equation of motion given as a connection of the dynamical
variables
\newpage

 along the elementary plaquette

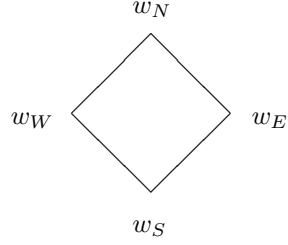
\begin{center}
\begin{picture}(120,110)(0,0)
  \multiput(30,60)(30,-30){2}{\line(1,1){30}}
  \multiput(30,60)(30,30){2}{\line(1,-1){30}}
  \put(7,56){$w_W$}   \put(53,98){$w_N$}
  \put(53,15){$w_S$}   \put(98,56){$w_E$}
\end{picture}
\end{center}

\[ w_N = f(qw_W) w_S (f(qw_E))^{-1} \;\; , \]
where
\[ f(w) = \frac{1+ \kappa^2 w}{\kappa^2 + w} \]
and $\kappa^2$ is a fixed real positive parameter. Function $f$ maps a unit
circle onto itself and is "odd"
\[ f(\frac{1}{w}) = (f(w))^{-1} .  \]

The same equation can be rewritten in terms of the dual vertex-like
variables $\psi$
\[ \psi_N = \psi_S f(q^{-1} \frac{\psi_W}{\psi_E}) . \]
To check it one is to multiply two last relations according to the picture for
two $\psi$-plaquettes

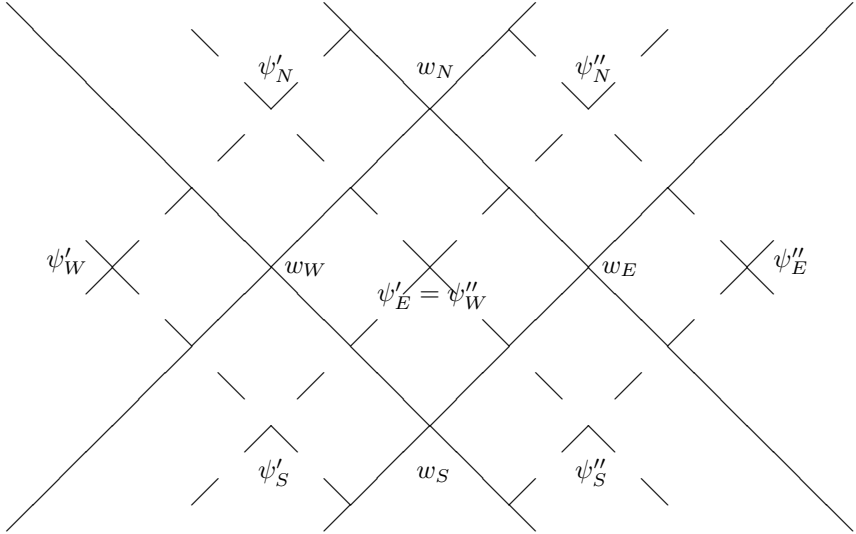
\begin{picture}(350,240)(0,40)
\multiput(40,160)(120,0){2}{\multiput(0,0)(20,20){5}{\line(1,1){10}}}
\multiput(160,160)(120,0){2}{\multiput(0,0)(-20,20){5}{\line(-1,1){10}}}
\multiput(160,160)(120,0){2}{\multiput(0,0)(-20,-20){5}{\line(-1,-1){10}}}
\multiput(40,160)(120,0){2}{\multiput(0,0)(20,-20){5}{\line(1,-1){10}}}

\multiput(100,160)(120,0){2}{\line(1,1){100}}
\multiput(100,160)(120,0){2}{\line(-1,1){100}}
\multiput(100,160)(120,0){2}{\line(1,-1){100}}
\multiput(100,160)(120,0){2}{\line(-1,-1){100}}

\put(40,160){\line(-1,1){10}}  \put(40,160){\line(-1,-1){10}}
\put(280,160){\line(1,1){10}}  \put(280,160){\line(1,-1){10}}

\put(15,160){$\psi_W^{\prime}$}     \put(95,233){$\psi_N^{\prime}$}
\put(95,80){$\psi_S^{\prime}$}      \put(215,80){$\psi_S^{\prime \prime}$}
\put(215,233){$\psi_N^{\prime \prime}$}
\put(290,160){$\psi_E^{\prime \prime}$}
\put(140,147){$\psi_E^{\prime}=\psi_W^{\prime \prime}$}
\put(105,158){$w_W$}    \put(225,158){$w_E$}
\put(155,233){$w_N$}    \put(155,80){$w_S$}
\end{picture}   \\
using the relation like
\[ w_W = \psi_N^{\prime \prime} (\psi_N^\prime)^{-1}  \]
and the commutation relations.

To get the classical continuous limit one must make one preliminary step --
a change of variables
\[ \chi = \left\{ \begin{array}{c} \psi \\
                                  \psi^{-1}  \end{array}
                                  \right.         \]

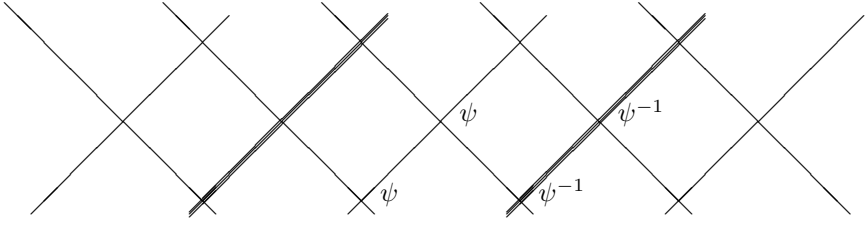
\begin{picture}(350,105)(0,20)
 \multiput(60,70)(60,0){5}{\line(1,1){40}}
 \multiput(60,70)(60,0){5}{\line(-1,1){45}}

 \multiput(60,70)(60,0){5}{\line(-1,-1){35}}     \put(187,70){$\psi$}
  \multiput(60,70)(60,0){5}{\line(1,-1){35}}     \put(157,40){$\psi$}

 \multiput(120,69)(0,2){2}{\line(1,1){40}}       \put(247,70){$\psi^{-1}$}
 \multiput(120,69)(0,2){2}{\line(-1,-1){35}}     \put(217,40){$\psi^{-1}$}

 \multiput(240,69)(0,2){2}{\line(1,1){40}}
 \multiput(240,69)(0,2){2}{\line(-1,-1){35}}

\end{picture}   \\
inverting $\psi$ on each second $SW - NE$ diagonal;
consistency with periodicity requires that $M$ is even.

The equation in new variables
\[ \chi_N = \chi_S^{-1} f(q^{-1} \chi_W \chi_E ) \]
turnes into the Sine-Gordon equation in the classical limit if we put
\[ \chi = e^{i \phi /2}   \]
(no rescaling for $\phi$!), and scale $\kappa^2$ in such a way
\[ \frac{1}{\Delta^2 \kappa^2} = \frac{m^2}{2}  . \]
The form of the lattice S-G equation
\[ \sin{\frac{1}{4} (\phi_W + \phi_E - \phi_N - \phi_S )} =
 \frac{1}{\kappa^2} \sin{\frac{1}{4} (\phi_W + \phi_E + \phi_N + \phi_S )}  \]
was first introduced in \cite{16} and revisited from the hamiltonian
point of view in \cite{17}, \cite{10}. However for our goal the form of
equation  in terms of $w$ (or $\psi$) is preferable, and we shall stick to
it, following  \cite{10}.

The lattice form of SG equation have the nice properties of its classical
continuous contraction, i.e., it can be written as a zero curvature
condition, has enough commuting conservation laws, etc.

To show it we introduce the appropriate Lax operator with 2-dimensional
auxiliary space. It plays the role of connection along the elementary
edge $B \leftarrow A$ of the lattice
\[ L_{B \leftarrow A} (\lambda) = \left( \begin{array}{cc}
 \psi_A^{-1/2} \psi_B^{1/2}    & \lambda \psi_A^{-1/2}  \psi_B^{-1/2}  \\
 \lambda \psi_A^{1/2} \psi_B^{1/2} & \psi_A^{1/2}  \psi_B^{-1/2}
 \end{array} \right) .          \]

One sees one more change of normalization $\psi \rightarrow \psi^{1/2}$,
but it seems to be inevitable. The equation of motion now takes the form
\[ L_{W \leftarrow N} (\kappa^{-1} \lambda ) L_{N \leftarrow E} (\kappa \lambda
) =
 L_{W \leftarrow S} (\kappa \lambda ) L_{S \leftarrow E} (\kappa^{-1} \lambda )
 \]
of a zero curvature along the elementary plaquette.

The conservation laws can be obtained by the traditional construction of
the transfer matrix;  however in our case one is to multiply the Lax
operator along the initial saw. With the notation
$L_{n-\frac{1}{2}} (\lambda)$  for the Lax operator connecting the
vertices with  $\psi_n$ and $\psi_{n-1}$ we have
\[ t(\lambda) = {\rm tr} (L_{2M-\frac{1}{2}} (\kappa \lambda)
L_{2M-\frac{3}{2}} (\kappa^{-1} \lambda)  \ldots  L_{\frac{3}{2}} (\kappa
\lambda)
L_{\frac{1}{2}} (\kappa^{-1} \lambda) C_2^{\frac{1}{2} \sigma_3} ) , \]
where the martrix $C_2^{\frac{1}{2} \sigma_3}$ takes care of the
periodicity.
The commutativity
\[   [ t(\lambda) , t(\mu) ] = 0   \]
is not evident because the Lax operators are not ultralocal; however the
ultralocality is restored when one observes that the composed Lax operator
\[  {\bf L}_n (\lambda) = L_{2n - \frac{1}{2}} (\kappa \lambda)
    L_{2n - \frac{3}{2}} (\kappa^{-1} \lambda)    \]
can be expressed through the canonical pair $w_{2n-1}, \psi_{2n-1}$
(generators of the algebra ${\cal C}_{odd}$)
\[ {\bf L}_n (\lambda)
 = \lambda \left(  \begin{array}{cc}
  \lambda^{-1}w^{\frac{1}{2}}_{2n-1} + \lambda w^{-\frac{1}{2}}_{2n-1}  &
 \psi^{\frac{1}{2}}_{2n-1} (\kappa w^{\frac{1}{2}}_{2n-1} + \kappa^{-1}
w^{-\frac{1}{2}}_{2n-1} ) \psi^{\frac{1}{2}}_{2n-1}   \\
 \psi^{-\frac{1}{2}}_{2n-1} (\kappa^{-1} w^{\frac{1}{2}}_{2n-1} + \kappa
w^{-\frac{1}{2}}_{2n-1} ) \psi^{-\frac{1}{2}}_{2n-1}   &
 \lambda w^{\frac{1}{2}}_{2n-1} + \lambda^{-1} w^{-\frac{1}{2}}_{2n-1}
 \end{array} \right)    \;\; . \]
 The quasiperiodicity of ${\bf L}_n (\lambda)$
\[ {\bf L}_{M+1} (\lambda) = C_2^{\frac{1}{2} \sigma_3} L_1 (\lambda)
 C_2^{-\frac{1}{2} \sigma_3}  \]
makes the apperance of $C_2^{\frac{1}{2} \sigma_3}$ in the definition of
$t(\lambda)$ quite natural.

The operator ${\bf L}_n (\lambda)$ is a variant of the lattice SG Lax
operator first introduced in \cite{18}. The posibility of its factorization
into the light-cone Lax operator with   the vertex-like dynamical variables
was realized  much later; it was indicated in \cite{19} and stressed in
\cite{17}.

The Lax operator ${\bf L}_n (\lambda)$ is ultralocal and satisfies the
ordinary fundamental commutation relation
\[ {\bf R} (\frac{\lambda}{\mu}) {\bf L}_n (\lambda) \otimes {\bf L}_n (\mu) =
 {\bf L}_n (\mu) \otimes {\bf L}_n (\lambda)  {\bf R} (\frac{\lambda}{\mu})
\]
 with the usual $4 \times 4$ trigonometric $R$-matrix. The commutativity
 of $t(\lambda)$ follows form this. Now $t(\lambda)$ is a polinomial in
$\lambda^2$ of degree $M$, and its coefficients are the independent
conservation laws in accordance with the number of degrees of freedom.

We can turn to the construction of the evolution operator, which generates
the main equation of motion. Some experimentation (or the didactic way,
first underlined in \cite{20}) leads to the Ansatz
\[ U = \prod r( \kappa^2, w_{even}) \prod r( \kappa^2, w_{odd})   \; ,   \]
where $r(\lambda, w)$ satisfies the functional equation
\[ \frac{r(\lambda, qw)}{r(\lambda, q^{-1}w)} = \frac{1+ \lambda w}{ \lambda +
w}  \;\;. \]
One can show that the map
\[ \psi \rightarrow \tilde{\psi} = U \psi U^{-1}  \]
is equivalent to the equation of motion directly by using the commutation
relations  between $\psi$-s and $w$-s and the functional equation.  Of course
the same is true if one choses $w$-s as the main  variables.
The functional equation is quite   inherent to our case. Indeed, it is
equivalent to the fundamental commutation relation
\[ r(\frac{\lambda}{\mu}, w_n) L_{n + \frac{1}{2}} (\lambda)   L_{n -
\frac{1}{2}} (\mu) =
  L_{n + \frac{1}{2}} (\mu) L_{n - \frac{1}{2}} (\lambda)
r(\frac{\lambda}{\mu}, w_n)  \; , \]
which is a variant of the fundamental commutation relation in the quantum
space  in sense of \cite{3}.

It is evident that
\[ r( \lambda, w) = \frac{S(w) S(w^{-1})}{S(\lambda w) S(\lambda w^{-1})}  \]
is a solution of the functional equation, if $S(w)$ satisfies the functional
equation of the previous  section. So one faces the alternative of referring
to compact or noncompact variant.

In the compact case one can treat the quantum lattice SG equation only for
rational coupling constant
\[ \gamma = \pi \frac{P}{Q} \;\;. \]
 The Hilbert space at the given site depends on $\gamma$ irregulary, its
 dimension being proportional to $Q$. In spite of the fact that $S(w)$
makes no sense for $|q| =1$, the ratio defining $r(\lambda, w)$ can be
truncated and leads to the expression for odd $Q$
\[ r(\lambda, w) = 1 + \sum^{\frac{Q-1}{2}}_{-\frac{Q-1}{2}}
 \frac{(1-\lambda)(q- \lambda q^{-1}) \cdots (q^{k-1} - \lambda q^{-k+1})}
 {(q^{-1} - \lambda q) \cdots (q^{-k} - \lambda q^k)} w^k    \]
first found in \cite{21}.

In the noncompact case everything depends on $\gamma$ smoothly, and for
$r(\lambda, w)$ we have the expression
\[ r(\lambda, w)  = \exp \frac{1}{2} \int^{\infty}_{\infty}
\frac{\cosh p \xi (1 - e^{-iw \xi})}{\sinh \pi \xi  \sinh \gamma \xi }
\frac{d \xi}{\xi}  \;\; , \]
where we put $ \lambda = e^w$, which irresistively reminds the Zamolodchikovs
phase-shift for the solutions ofthe SG model \cite{22}. Needless to
stress that in our treatment it appeares in a completely different function
-- as a local factor, defining the evolution operator, the variable $p$
(rapidity in the phase-shift interpretation) being a local current-like
dynamical variable. The origine and the interpretation of this duality
is still      unclear but work on this direction is most promising.

Let us repeat, that with respect to dynamics both languages reconcile, if
we consider the noncompact $U$ as a outer automorphism of the algebra
$\cal A$. However, the Hilbert space interpretation of the algebraic
formulas is rather different for compact or noncompact choice. This
difference seems rather irrelevant in the continuous limit, which is known
to exist in quantum case if accompanied with the nontrivial renormalization
of the parameter $\kappa$.

Now we are in a position to make the comparison of massive and massless case.

\section{Discussion and conclusions}

In the main body of these lectures I have shown that one can get a rather
unified treatment of the massive and massless intrgrable models on the
lattice,
using the current-like dynamical variables $w$. In particular, one gets the
possibility to compare both cases on the same common ground.

The massless contraction is rather evident now. The equation of motion
\[ w_N = w_W w_S w_E^{-1}  \;  , \]
which can be obtained from the one in section 4 in the limit $\kappa^2
\rightarrow \infty $ clearly separates: the variables
\[ \xi_n = w_{2n} w_{2n +1}^{-1}  \]
and
\[ \eta_n = w_{2n} w_{2n +1}  \]
propagate independently by means of the left or right shift  correspondingly.
These variables are also current-like
\[ \xi_{n+1} \xi_n = q^2  \xi_n \xi_{n+1}     \]
\[ \eta_n \eta_{n+1} = q^2 \eta_{n+1} \eta_n   \]
and mutually commute
\[ \xi_n \eta_m = \eta_m \xi_n  \; . \]
The separation in the evolution
\[ U_{massive}(w) |_{\kappa^2 \rightarrow \infty } = U_{massless} (\xi)
 U_{massless} (\eta)       \]
 is based on the formula
\[ r(\kappa^2, w) |_{\kappa^2 = \infty} = r(w)   \]
and commutation relations
\[ r(w_n) r(w_{n+1})  = r(w_{n+1}) r(q^{-1}w_{n+1}^{-1}w_n)   \]
\[ r(w_{n+1}) r(w_n)  = r(w_n) r(q^{-1}w_n w_{n+1})   \]
which follows from the functional equations for $r(w)$ and commutation
relations  for $w_n$.

An interesting question appears on this level: what remains of the two
lattice Virasoro algebras of massless case in the massive deformation
(or more constructively, can they be combined into some new algebraic
structure)? This question is not answered yet and deserves more work.

Less superficial results on the interrelation of massive and massless cases
appear on the level of the Bethe-Ansatz equations; we cannot discuss them
here, as we did not mention Bethe-Ansatz at all. Work in this direction
is quite active. In particular the use of Bethe-Ansatz together with
the comments on duality in the section 3 must throw a new light on the
appearence of traces of two quantum groups in Kac table. The minimal
models are to be obtained as massless contraction for special rational
$\gamma$
\[ \gamma = \pi \frac{p}{p+1}  \;\; .\]

Besides the field-theoretical applications the considerations in these
lectures are relevant to some pure mathematical questions. We mention
two of them.

1. Quite a few formulas here resemble the treatment of the algebra
$U_q(\hat{sl(2)})$ in \cite{23}. The etablishing more close connection
is certainly desirable.

2. Classical limit $\hbar \rightarrow 0$ retaining finite lattice site
$\Delta$ give us a family of models for the integrable symplectic maps.
As such it is a valuable addition to the existing results in this field
\cite{24}, \cite{25}.

In these lectures we have treated only the simplest case of the abelian
current-like variables. The nonabelian analogue -- lattice Kac-Moody
algebra -- is also known \cite{26} \cite{27}. It is clearly relevant
to the lattice WZNW model, which was discussed in \cite{27}.

However some principal developments, i.e., the Sugawara construction
of Virasoro algebra, are still not finished and work is in progress.

\end{document}